# Mode analysis of a class of spatiotemporal photonic crystals

**Juan C. González[1], Juan C. Miñano [1,2] and Pablo Benítez [1,2]**

[1]*Cedint. Universidad Politécnica de Madrid. Campus de Montegancedo. 28223. Madrid. Spain*

[2] *LPI, 2400 Lincoln Avenue, Altadena, CA 91001, USA*

<u>jcgonzalez@cedint.upm.es, jc.minano@upm.es, pbenitez@etsit.upm.es</u>,

**Abstract**

A detailed analysis is presented of the modes of a class of spatiotemporal photonic crystal. The structure analyzed is a perfect dielectric with periodic variation of $\varepsilon$ in a single spatial direction, as well as periodic variation in time. Analytic solutions are presented for $\varepsilon$ being separated in both variables, that is $\varepsilon(z,t)=\varepsilon_0\varepsilon_r(z)\varepsilon_t(t)$, and dispersion diagrams are presented.

## 1. Introduction

The interaction between EM waves at optical frequencies and materials with time-varying properties has been broadly analyzed. Filtering, switching, wavelength conversion, light capture, and waveguide or optical amplification have been analyzed as possible applications of these systems [1-5]. Phononic crystals with time varying parameters are another field of development with applications similar to photonic crystals [6,7]. For periodic $\varepsilon(\mathbf{r},t)$ in $\mathbf{r}$ or $t$, Bloch's theorem can be applied and the dispersion diagram is obtained by solving linear systems of equations [8,9]. Electric ($\mathbf{E}(\mathbf{r},t)$) and Magnetic ($\mathbf{H}(\mathbf{r},t)$) fields are expressed as modulated periodic functions. Analytical, numerical and semi-numerical methods are also used [10]. Dielectric structures with dielectric constant $\varepsilon$ non-depending on time and spatially periodic are well known ($\varepsilon(\mathbf{r},t)=\varepsilon(z)=\varepsilon(z+L)$ for 1D PhC) . Structures with non-spatial dependence and periodic dependence on time can be analyzed with the same procedure than 1D spatial PhC ($\varepsilon(\mathbf{r},t)=\varepsilon(t)=\varepsilon(t+T)$) as it will be shown in the next paragraph. Finally, structures with spatiotemporal dependence, $\varepsilon(\mathbf{r},t)=\varepsilon(p)=\varepsilon(p+\Lambda_p)$ with $p=k_o z+\omega_o t$, has been also broadly studied. These three particular types of structures present similar dispersion diagrams because the Bloch functions used are periodic in only one variable, *z* for 1D spatial PhC, *t* for temporal PhC and *p* for the third type mentioned. The spatiotemporal PhC analyzed in these papers presents periodic dependence in *z* and *t*, that is, $\varepsilon(\mathbf{r},t)=\varepsilon(z,t)=\varepsilon(z+L,t+T)$. Bloch functions in these structures are periodic in both variables, and dispersion diagrams have characteristics completely different. The analysis can be greatly simplified if the dielectric constant (for simplicity we are going to assume $\mu=\mu_o$=constant) can be separated in its spatial and temporal variables, that is $\varepsilon(\mathbf{r},t)=\varepsilon_o\varepsilon_r(\mathbf{r})\varepsilon_t(t)$. In systems with these characteristics, separation of variables methods can be applied to obtain analytical solutions. This procedure has been used in analysis of Bragg reflectors with time varying dielectric constants [11]. In this paper we present a method of finding analytical solutions of the modes in 1D Photonic Crystals (PhC) with separately periodic space-time varying $\varepsilon$, where the separated field $\mathbf{H}(\mathbf{r},t)=\mathbf{H_r}(\mathbf{r})H_t(t)$ can be obtained with Bloch theorem. The relation between the **k** and $\omega$ parameters of the Bloch function of $\mathbf{H_r}(\mathbf{r})$ and $H_t(t)$ (dispersion diagram) is expressed in parametric form through one real parameter *P* ( **k**(*P*) and $\omega(P)$). Each mode is defined as a solution of the field for a particular pair of the parameters **k** and $\omega$.

## 2. Analogy between one dimensional spatial, temporal and spatiotemporal photonic crystal.

One dimensional (1D) photonics crystals with periodic spatial variation of dielectric constant $\varepsilon$ have been broadly studied [12]. In these systems, $\varepsilon$ is periodic in *z*, that is, $\varepsilon(z)=\varepsilon(z+L)=\varepsilon_0\varepsilon_r(z)$ ($\varepsilon_0$, dielectric constant in vacuum, $\varepsilon_r$, relative dielectric constant). Uniform media with temporal variation of $\varepsilon$ ($\varepsilon(t)=\varepsilon(t+T)=\varepsilon_0\varepsilon_t(t)$) can be analyzed in a similar way. The equation for the **H** field in a 1D spatial crystal (1) and temporal crystal (2) are obtained directly from Maxwell's equations:

$$\nabla x(\frac{1}{\varepsilon_r(z)}\nabla x\mathbf{H}_{0r}(\mathbf{r}))-\left(\frac{\omega}{c}\right)^2\mathbf{H}_{0r}(\mathbf{r})=0$$
$$\mathbf{H}_r(\mathbf{r},t)=\mathbf{H}_{0r}(\mathbf{r})\exp(-j\omega t) \quad \mathbf{E}_r(\mathbf{r},t)=\mathbf{E}_{0r}(\mathbf{r})\exp(-j\omega t)$$
(1)

$$-k^2\mathbf{H}_{0t}(t)-\frac{1}{c^2}\frac{\partial}{\partial t}(\varepsilon_r(t)\frac{\partial \mathbf{H}_{0t}(t)}{\partial t})=0$$
$$\mathbf{H}_t(\mathbf{r},t)=\mathbf{H}_{0t}(t)\exp(-j\mathbf{kr}) \quad \mathbf{E}_t(\mathbf{r},t)=\mathbf{E}_{0t}(t)\exp(-j\mathbf{kr})$$
(2)

Where $c$ is the speed of light in vacuum. Due the periodicity of $\varepsilon$ with $z$ or $t$, Bloch's theorem can be used in both crystals, so that the modes are (3):

$$\mathbf{H}_{\mathbf{rk}\omega}(r,t)=\exp[j(\mathbf{kr}-\omega t)]\mathbf{u}_{\mathbf{rk}\omega}(z) \quad \mathbf{H}_{\mathbf{t}\omega\mathbf{k}}(\mathbf{r},t)=\exp[j(\mathbf{kr}-\omega t)]\mathbf{u}_{\mathbf{t}\omega\mathbf{k}}(t)$$
$$\mathbf{H}_{\mathbf{rk}\omega}(r,t)=\exp[j(\mathbf{kr}-\omega t)]\sum_n\mathbf{H}_{\mathbf{zn}}\exp(jnk_pz) \quad k_p=\frac{2\pi}{L}$$
$$\mathbf{H}_{\mathbf{t}\omega\mathbf{k}}(\mathbf{r},t)=\exp[j(\mathbf{kr}-\omega t)]\sum_n\mathbf{H}_{\mathbf{tn}}\exp(jn\omega_pt) \quad \omega_p=\frac{2\pi}{T}$$
(3)

where $\mathbf{u}_{\mathbf{rk}\omega}(z)$ and $\mathbf{u}_{\mathbf{t}\omega\mathbf{k}}(t)$ are periodic functions with periods $L$ and $T$, respectively. For the spatial crystal, $\omega$ is real and **k** is a complex function of $\omega$, $\mathbf{k}(\omega)=\mathbf{k}_r(\omega)+j\mathbf{k}_i(\omega)$, which is obtained from the periodicity condition of $\mathbf{u}_{\mathbf{rk}\omega}(z)$. For the temporal crystal, $\omega$ is a complex function of $k$, $\omega(k)=\omega_r(k)+j\omega_i(k)$ ($k$ is the modulus of **k** that is real in temporal crystal). If the change $\mathbf{k'}=\mathbf{k}+n_0k_p\mathbf{z}$ is made in the spatial crystal the expression of the field with this new wavevector is (4):

$$\mathbf{H}_{rk\omega}(r,t) = \exp[j(\mathbf{k'r} - \omega't)]\sum_n \mathbf{H'}_{zn}\exp(jnk_p z) =$$
$$\exp[j(\mathbf{kr} - \omega't)]\sum_n \mathbf{H'}_{zn}\exp[j(n+n_o)k_p z] \qquad (4)$$

This expression is identical to (3) with the condition $\omega'=\omega$ and $\mathbf{H'}_{zn}=\mathbf{H}_{z(n+no)}$. Due to this property the dispersion diagram is periodic in the variable $k$. Fig.1 shows a typical diagram of the functions $k_r(\omega)$ and $k_i(\omega)$ for TEM solutions, $\mathbf{k}=k\mathbf{z}$, of a spatial crystal (functions are symmetrical with respect to both axes). The function $k(\omega)$ has the following characteristics:

1. The inverse function of $k_r(\omega)$, i.e., $\omega(k_r)$ is periodic with period $2\pi/L$. This function is also multi-valued, for each $k_r$, there are an infinite number of possible values of $\omega$.

2. Intervals in $\omega$ where $k_r(\omega)\neq n\pi/L$ (n is an integer) and $k_i(\omega)=0$ are called transmitted modes.

3. Intervals in $\omega$ that fulfill $k_r(\omega)=n\pi/L$ and $k_i(\omega)\neq 0$ correspond to untransmitted modes. The field $\mathbf{H}_{rk\omega}$ in these intervals has an exponential envelope, as can be seen in Eq. (2).

Due to the periodicity of $\omega(k_r)$, the representation in the interval $[0, \pi/L]$ of $k_r$ (Fig.1, solid lines) gives all information about the function $\omega(k_r)$. An alternative and equivalent representation of $\omega(k_r)$ is also shown in Fig.1 (bold lines). The function is not restricted to a single period in the variable $k_r$ but it is a single-valued function. The curves are similar to those of Fig.1 for temporal PhC. It is only necessary to replace $k$ by $\omega$, $\omega_r$ by $k_r$ and $\omega_i$ by $k_i$. Fig. 2 shows the function $\omega_r(k)$ with $\omega_r$ in horizontal and vertical axis. It is clear from Fig. 2 that the expression $v_g=d\omega_r/dk$ is higher than speed of light in vacuum for some intervals in $\omega_r$. This superluminic phenomenon has been broadly analyzed in both absorbing and amplifying media [13-15] and its detailed analysis is beyond the scope of this paper. For intervals of $\omega_i \neq 0$, the fields are exponentially decaying or growing vs time, i.e., the fields are attenuated or amplified.

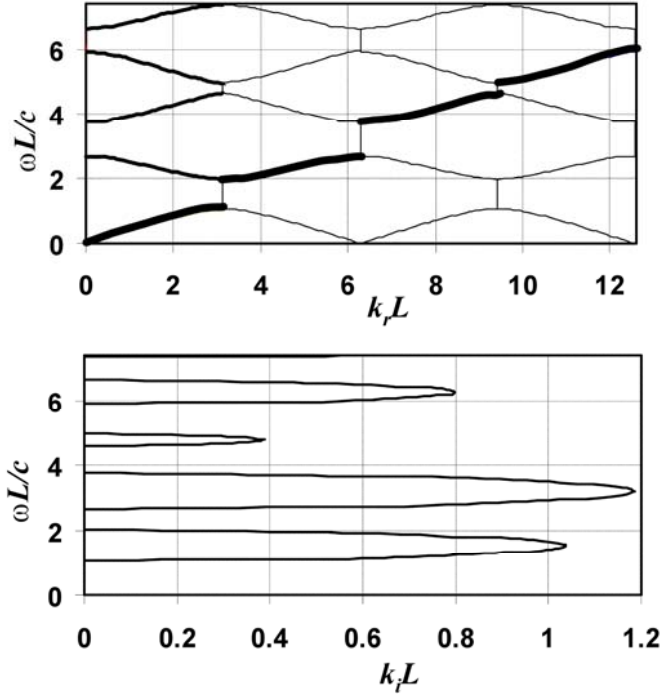

**Fig. 1.** (Up), real part of $k$ (horizontal axis) as a function of the real variable $\omega$ (vertical axis). Bold, solid and thin lines represent all the solutions. The solid lines represent the multi-valued function in one period. The bold lines are an alternative representation, complete axis $k_r$ and one-valued function. Vertical right lines, intervals with $k_i \neq 0$. The structure is stepped-index: $\varepsilon_r=1$ for $0<z<3$ and $\varepsilon_r=12$ for $3<z<5$. ($z$ in μm), $L=5$ μm.

(Down), imaginary part of $k$ (horizontal axis) as function of real variable $\omega$ (vertical axis).

Another family of structures related to Optical Amplifier [1] or moving media [16] is the dielectric variation (5):

$$\varepsilon(z,t) = \varepsilon(p) = \varepsilon(p+\Lambda_p)$$
$$p = \omega_0 t + k_0 z \tag{5}$$

Where $\omega_o$, $k_o$ and $\Lambda_p$ are constants. The analysis can be made using Bloch functions (6).

$$\mathbf{H}_{p\mathbf{k}\omega}(z,t) = \exp[j(\omega t + kz)]\mathbf{U}(p) = \exp[j(\omega t + kz)]\sum_n \mathbf{H}_{\mathbf{pn}} \exp(jn\frac{2\pi}{\Lambda_p}p) =$$

$$\exp[j(\omega t + kz)]\sum_n \mathbf{H}_{\mathbf{pn}} \exp[jn\frac{2\pi}{\Lambda_p}(\omega_o t + k_o z)] \tag{6}$$

$$\mathbf{U}(p) = \mathbf{U}(p+\Lambda_p)$$

As it happens in spatial and temporal crystals, modes (6) are solution of the equations if there is a relation between variables $\omega$ and $k$, that is, the dispersion diagram of this structure.

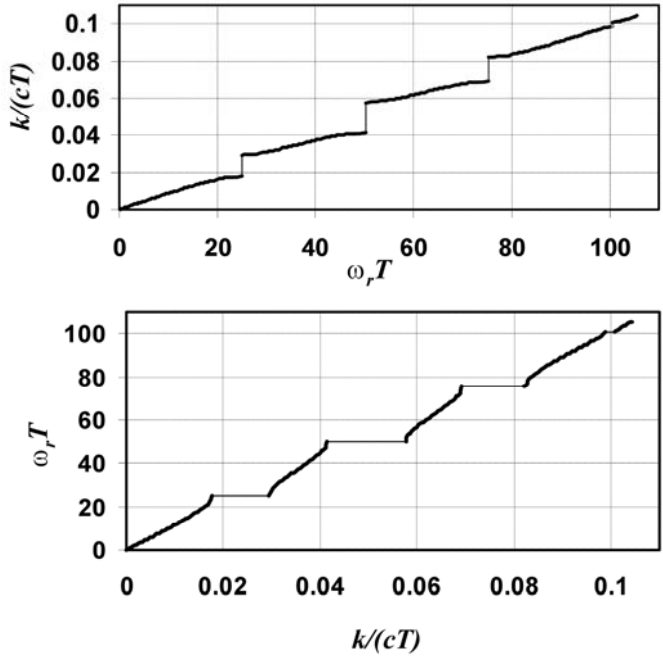

**Fig. 2.** The function $\omega_r(k)$ in temporal photonic crystals. The structure being analyzed is uniform in space and stepped-index in time: $\varepsilon_r=1$ for $0<t<5$ and $\varepsilon_r=10$ for $5<t<8$. ($t$ in $10^{-15}$s), T=$5 \cdot 10^{-15}$s. In the first representation (up), $\omega_r$ in horizontal axis, the untransmitted bands are the vertical lines. With $\omega_r$ on the vertical axis the untransmitted modes are the horizontal lines (down).

For frequency $\omega'=\omega+n_o\omega_o$ and wavenumber $k'=k+n_o k_o$, the Bloch functions are (7):

$$\mathbf{H}_{pk\omega}(z,t) = \exp[j(\omega't+k'z)]\mathrm{U}'(p) = \\ \exp[j(\omega t+kz)]\sum_n \mathrm{H'}_{pn}\exp[j(n+n_o)\frac{2\pi}{\Lambda_p}(\omega_o t+k_o z)] \quad (7)$$

According to (6) and (7) if $(k, \omega)$ is a point of the dispersion diagram, $(k+n_o k_o, \omega+n_o\omega_o)$ is also a point of the dispersion diagram and:

$$\mathrm{H'}_{pn} = \mathrm{H}_{p(n+n_o)} \quad (8)$$

So the dispersion diagram is similar to the spatial PhC but rotating the $k$ axis up to the right (Fig. 3):

$$\omega = k\frac{\omega_o}{k_o} \quad (9)$$

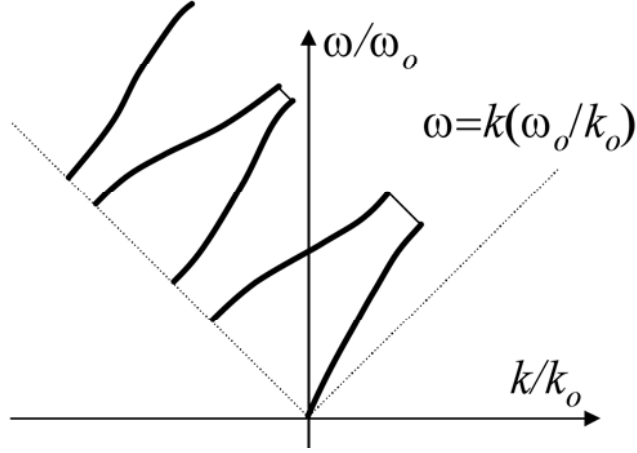

**Fig. 3.** One period of the dispersion diagram of one structures of the type $\varepsilon(z,t)=\varepsilon(p)=\varepsilon(p+\Lambda_p)$ where $p=k_o z+\omega_o t$. In these systems the Bloch functions are of one dimensional expansion type and the dispersion diagram have similar characteristics than the spatial crystals. The curves are periodic in the direction $\omega=k(\omega_o/k_o)$. When $\omega_o=0$, spatial crystal is obtained (periodic in $k$ variable).

This ($k,\omega$) representation has been shown for example in Cassedy [9] and Biancalana [16]. It is important to note that the summation in (7), as happens in (4), is done on only one integer n, that is, both are one-dimensional expansions. The dispersion diagram is periodic in direction (9) but not in the perpendicular direction to this line.

A different type of structures are those with periodic $\varepsilon(z,t)$ in both variables and in which $z$ and $t$ are not related, that is, $\varepsilon(z, t) =\varepsilon(z+L, t+T)$. The Bloch function in this case is (10):

$$\begin{aligned}\mathbf{H}_{ztk\omega}(z,t) &= \exp[j(\omega t + kz)]\mathbf{U}(z,t) = \\ &\exp[j(\omega t + kz)]\sum_{n,m}\mathbf{H}_{n,m}\exp(jnk_o z)\exp(jm\omega_o t)\end{aligned} \quad (10)$$

$$\mathbf{U}(z,t) = \mathbf{U}(z+L,t+T) \qquad k_o = \frac{2\pi}{L} \qquad \omega_o = \frac{2\pi}{T}$$

Where now the summation is on two integers n and m, that is, it is a two dimensional expansion. For frequency $\omega'$ and wave number $k'$ such that (11):

$$k' = k + n_1 k_o \qquad \omega' = \omega + n_2 \omega_o \qquad (11)$$

the Bloch function are (12):

$$\mathbf{H}_{ztk\omega}(z,t) = \exp[j(\omega' t + k' z)]\mathbf{U}'(z,t) = \\ \exp[j(\omega t + kz)]\sum_{n,m}\mathbf{H'}_{n,m} \exp[j(n+n_1)k_o z]\exp[j(m+n_2)\omega_o z] \qquad (12)$$

So, if the pair ($\omega$, $k$) is a point of the dispersion diagram, point ($k+n_1k_o$, $\omega+n_2\omega_o$) is also a point of the diagram. In these structures, the dispersion diagram is then periodic in both axes, $k$ and $\omega$. To investigate the properties of this diagrams, structures with this type of dependence have been analyzed in the following parts of these papers but with the restriction of separated constant $\varepsilon$ ($\varepsilon(z,t)=\varepsilon_0\varepsilon_r(z)\varepsilon_t(t)$). Analytic solutions are found for any variation of $\varepsilon_r(z)$ and $\varepsilon_t(t)$. For the more general variation $\varepsilon(z, t) = \varepsilon(z+L, t+T)$ in which $\varepsilon$ is not separated, numerical methods are mandatory.

## 3. General solutions of a one-dimensional spatiotemporal photonic crystal for separated dielectric constant.

To solve Maxwell's equations with a dielectric constant periodic in $z$ and $t$, the separation of variables method is used (13):

$$\mathbf{H}(z,t) = \mathbf{H}_\mathbf{r}(z)H_t(t) = (H_{rx}(z)\hat{x} + H_{ry}(z)\hat{y} + H_{rz}(z)\hat{z})H_t(t) \\ \mathbf{E}(z,t) = \mathbf{E}_\mathbf{r}(z)E_t(t) = (E_{rx}(z)\hat{x} + E_{ry}(z)\hat{y} + E_{rz}(z)\hat{z})E_t(t) \qquad (13)$$

Introducing the fields of Eq. 13 into Maxwell's equations yields the following scalar equations for **H** (14, 15):

$$\frac{d}{dt}\left(\varepsilon_t(t)\frac{dH_t(t)}{dt}\right) + P^2 H_t(t) = 0 \qquad (14)$$

$$\frac{d}{dz}\left(\frac{1}{\varepsilon_r(z)}\frac{dH_{ry}(z)}{dz}\right)+\left(\frac{P}{c}\right)^2 H_{ry}(z)=0$$

$$\frac{d}{dz}\left(\frac{1}{\varepsilon_r(z)}\frac{dH_{rx}(z)}{dz}\right)+\left(\frac{P}{c}\right)^2 H_{rx}(z)=0 \quad (15)$$

$$H_{rz}(z)=0$$

Where $P^2$ is a constant (independent of $z$ or $t$) appearing in the classical separation of variables method. Bloch functions can be used to find $\mathbf{H_r}(z)$ and $H_t(t)$ from Equ. 14 and 15 (16, 17):

$$H_{rx}(z)=\exp(jkz)u_k(z)$$

$$\frac{d\left(\frac{1}{\varepsilon_r(z)}\frac{du_k(z)}{dz}\right)}{dz}+jk\frac{d\left(\frac{u_k(z)}{\varepsilon_r(z)}\right)}{dz}+jk\frac{1}{\varepsilon_r(z)}\frac{du_k(z)}{dz}+u_k(z)\left(\frac{(jk)^2}{\varepsilon_r(z)}+(\frac{P}{c})^2\right)=0 \quad (16)$$

$$H_t(t)=\exp(j\omega t)u_\omega(t)$$

$$j\omega\frac{d(\varepsilon_t u_\omega(t))}{dt}+(j\omega)^2\varepsilon_t u_\omega(t)+j\omega\varepsilon_t\frac{du_\omega(t)}{dt}+\frac{d\left(\varepsilon_t\frac{du_\omega(t)}{dt}\right)}{dt}+P^2 u_\omega(t)=0 \quad (17)$$

The function $u_k(z)$ is periodic in $z$ with period $L$, and $u_\omega(t)$ is periodic in $t$ with period $T$. Eqns. 15 and 16 are linear and homogeneous, so the general solution depends on two arbitrary constants and can be expressed as (18):

$$u_k(A,B,z,P,jk)=A\left[f_1(z,P,jk)+Bf_2(z,P,jk)\right]$$
$$u_\omega(C,D,z,P,jk)=C\left[g_1(z,P,j\omega)+Dg_2(z,P,j\omega)\right] \quad (18)$$

Where $f_1$ and $f_2$ are two particular linear independent solutions of Eq. 16, $g_1$ and $g_2$ are two particular linear independent solutions of Eq. 17, and $A$, $B$, $C$ and $D$ are arbitrary constants. The periodicity of the functions $u_k(z)$ and $u_\omega(t)$ are assured if the functions and their first derivatives are the same at the planes $z=0$, $z=L$ for $u_k(z)$, and at the instants $t=0$, $t=T$ for $u_\omega(t)$ (19, 20):

$$f_1(0,P,jk) + Bf_2(0,P,jk) = f_1(L,P,jk) + Bf_2(L,P,jk)$$
$$\left.\frac{df_1}{dz}\right|_{z=0} + B\left.\frac{df_2}{dz}\right|_{z=0} = \left.\frac{df_1}{dz}\right|_{z=L} + B\left.\frac{df_2}{dz}\right|_{z=L} \tag{19}$$

$$g_1(0,P,j\omega) + Dg_2(0,P,j\omega) = g_1(T,P,j\omega) + Dg_2(T,P,j\omega)$$
$$\left.\frac{dg_1}{dt}\right|_{t=0} + D\left.\frac{dg_2}{dt}\right|_{t=0} = \left.\frac{dg_1}{dt}\right|_{t=T} + D\left.\frac{dg_2}{dt}\right|_{t=T} \tag{20}$$

Relations $k=k(P)$, $B=B(P)$ and $\omega=\omega(P)$, $D=D(P)$ are obtained from Eq. 19 and 20. For each value of $k$ and $\omega$, there is one solution $\mathbf{H}_{k\omega}$ that is called the mode. The functions $k(P)$ and $\omega(P)$ are complex functions of the real variable $P$.

### 4. Example of a structure with ε stepped in $z$ and $t$.

Next is a complete analysis of one particular example. The dielectric constant is periodic and has the following expression for each period (21):

$$\varepsilon_r(z) = \begin{cases} \varepsilon_{z1} & 0 < z < z_1 \\ \varepsilon_{z2} & z_1 < z < L \end{cases} \qquad \varepsilon_t(t) = \begin{cases} \varepsilon_{t1} & 0 < t < t_1 \\ \varepsilon_{t2} & t_1 < t < T \end{cases} \tag{21}$$

Here $\varepsilon_{z1}$, $\varepsilon_{z2}$, $\varepsilon_{t1}$ and $\varepsilon_{t2}$ are constants. From Eq. 16 and 17, the functions $u_k(z)$ and $u_\omega(t)$ are:

$$u_k(z) = \begin{cases} a_1 \exp[(-jk+jk_1)z] + a_2 \exp[(-jk-jk_1)z] & 0 < z < z_1 \\ a_3 \exp[(-jk+jk_2)z] + a_4 \exp[(-jk-jk_2)z] & z_1 < z < z_2 \end{cases}$$
$$k_1 = \frac{P}{c}\sqrt{\varepsilon_{z1}} \qquad k_2 = \frac{P}{c}\sqrt{\varepsilon_{z2}} \tag{22}$$

$$u_\omega(z) = \begin{cases} b_1 \exp[(-j\omega + j\omega_1)z] + b_2 \exp[(-j\omega - j\omega_1)t] & 0 < t < t_1 \\ b_3 \exp[(-j\omega + j\omega_2)t] + b_4 \exp[(-j\omega - j\omega_2)t] & t_1 < t < T \end{cases} \quad (23)$$

$$\omega_1 = \frac{P}{\sqrt{\varepsilon_{t1}}} \qquad \omega_2 = \frac{P}{\sqrt{\varepsilon_{t2}}}$$

The contour conditions at $z=z_1$ and $t=t_1$, are derived [10,13]:

$$H_{rx}(z_1^-) = H_{rx}(z_1^+) \qquad \left(\frac{1}{\varepsilon_{z_1}}\frac{dH_{rx}(z)}{dz}\right)_{z=z_1^-} = \left(\frac{1}{\varepsilon_{z_2}}\frac{dH_{rx}(z)}{dz}\right)_{z=z_1^+} \quad (24)$$

$$H_t(t_1^-) = H_t(t_1^+) \qquad \left(\varepsilon_{t_1}\frac{dH_t(t)}{dt}\right)_{t=t_1^-} = \left(\varepsilon_{t_2}\frac{dH_t(t)}{dt}\right)_{t=t_1^+} \quad (25)$$

The constants $a_3$ and $a_4$ are calculated from $a_1$ and $a_2$ and the constants $b_3$ and $b_4$ from $b_1$ and $b_2$ with the conditions in $z_1$ for Eq. 24 and in $t_1$ for Eq. 25. The periodicity condition is obtained with Eq. 24 and Eq.25 in $z=L$ and $t=T$ and Eq. 19 and 20. Fig.4 and 5 show the real and imaginary parts of $k$ and $\omega$ as functions of P for one example.

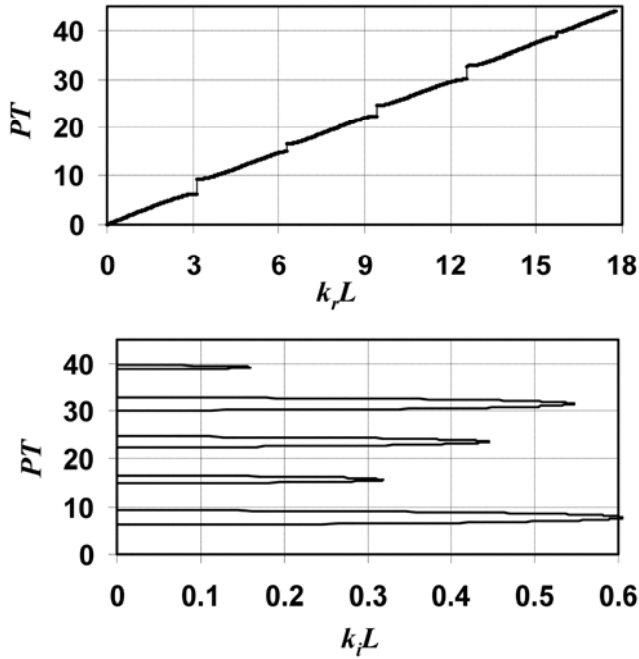

**Fig. 4.** Real (up) and imaginary (down) part of k (horizontal axis) as a function of P (vertical axis). The dimensions of *P*, from Eq. 14 and 15, are the same as $\omega$ so *PT* has a non-dimensional magnitude. The spatial variation of the index (Eq. 21) is: $\varepsilon_{z1}=1$, $\varepsilon_{z2}=9$, $z_1=3$ μm and $z_2=5$ μm.

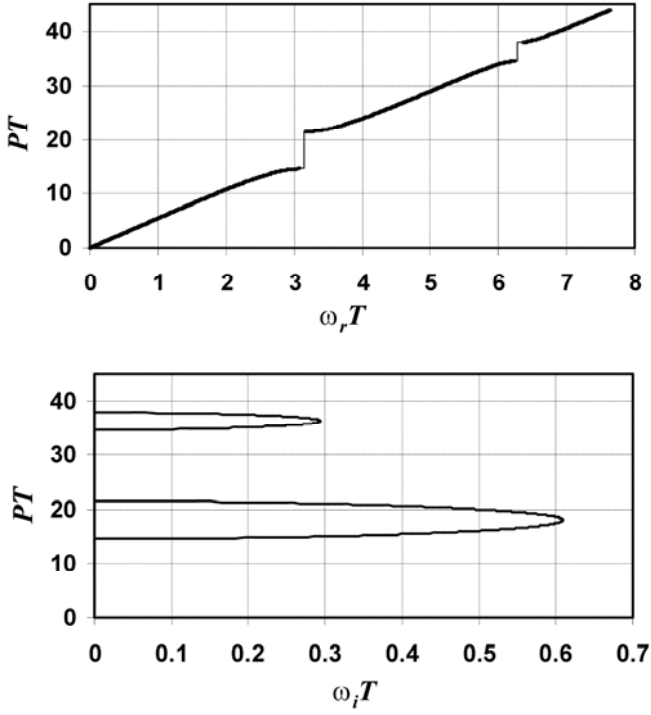

**Fig. 5.** Real (up) and imaginary (down) part of $\omega$ as function of $P$. The temporal variation of index is: $\varepsilon_{t1}=1$, $\varepsilon_{t2}=10$, $t_1=3\cdot 10^{-15}$ s and $t_2=8\cdot 10^{-15}$ s.

These functions fulfill similar characteristics to the function $k(\omega)$ of a spatial Photonic Crystal and $\omega(k)$ of a temporal one:

1. The functions $P(\omega_r)$ and $P(k_r)$ are periodic with period $2\pi/L$ and $2\pi/T$. These functions are multi-valued: for each $k_r$ and $\omega_r$ there are an infinite number of solutions $P$.

2. Intervals in $P$ where $k_r(P)\neq n\pi/L$ (n is an integer), $k_i(P)=0$ and $\omega_r(P)\neq m\pi/T$, $\omega_i(P)=0$ correspond to transmitted modes.

3. Intervals in $P$ where $k_r(\omega)=n\pi/L$, $k_i(\omega)\neq 0$ correspond to untransmitted modes in $z$. The field $\mathbf{H}_{k\omega}$ has an exponential envelope in variable $z$.

4. Intervals in $P$ where $\omega_r(P)=m\pi/T$, $\omega_i(P)\neq 0$ correspond to untransmitted modes in $t$. The field $\mathbf{H}_{k\omega}$ has an exponential envelope in variable $t$.

Fig.6 shows the dispersion diagram thus obtained for the curves of Fig.4 and 5. $\omega_r T$ is represented in the vertical axis and $k_r L$ in the horizontal. The vertical lines correspond to a mode that is disallowed due to $k_i\neq 0$, and the horizontal ones for modes disallowed due to $\omega_i\neq 0$.

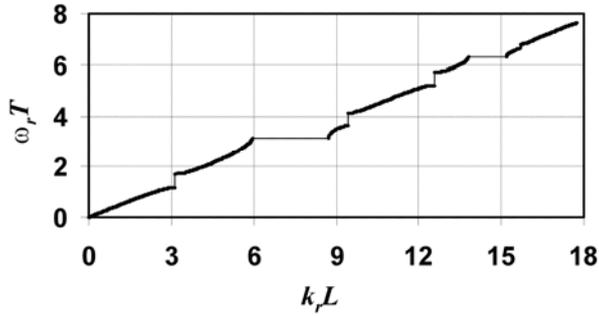

**Fig. 6.** Dispersion diagram. This representation is the relation between real parts of ω and k. Horizontal and vertical right lines are the intervals of not existence of modes.

## 6. Conclusions.

An analytical procedure has been applied to spatiotemporal photonic crystals. The unique condition of the system is the separation of the dielectric constant as the product of spatial and temporal functions. A dispersion diagram is presented in parametric form through an intermediate variable *P*. A classical dispersion diagram $\omega(k)$ can be obtained from the parametric form and can be interpreted similarly. For systems of stepped index in both variables, analytical solutions are obtained and other index variations can be modeled with stepped functions using the same analysis.

## 7. Acknowledgement.

The authors thank the Spanish Ministries MCEI (Consolider program CSD2008-00066, DEFFIO: TEC2008-03773), MITYC (OSV: TSI-02303-2008-52), and the Madrid Regional Government (LED-TV: 130/2008 TIC, ABL: PIE/466/2009, F3: PIE/469/2009 and CAM/UPM-145/Q060910-103) for the support given in the preparation of the present work.

## 8 References.